\documentclass[conference]{IEEEtran}
\IEEEoverridecommandlockouts
\usepackage{cite}
\usepackage{amsmath,amssymb,amsfonts}
\usepackage{algorithmic}
\usepackage{graphicx}
\usepackage{textcomp}
\usepackage{xcolor}
\usepackage{hyperref}
\def\BibTeX{{\rm B\kern-.05em{\sc i\kern-.025em b}\kern-.08em
    T\kern-.1667em\lower.7ex\hbox{E}\kern-.125emX}}
    
\usepackage{enumitem}

\usepackage{microtype}
\usepackage{xcolor}
\usepackage{graphicx}
\usepackage{subfigure}
\usepackage{booktabs} 
\usepackage{enumitem}
\usepackage{pifont}
\usepackage{multirow}
\usepackage{amssymb}

\newlist{myitemize}{enumerate}{10}
\setlist[myitemize]{label*=\arabic*.,leftmargin=0pt}

\usepackage{hyperref}

\title{Voice Reenactment with F0 and timing constraints \\ and adversarial learning of conversions}

\author{\IEEEauthorblockN{Frederik Bous, Laurent Benaroya, Nicolas Obin, Axel Roebel}
\IEEEauthorblockA{STMS Lab  \\
IRCAM, CNRS, Sorbonne Université \\
Paris, France}}

\begin{document}
%
\maketitle
\begin{abstract}
This paper introduces {\em voice reenactement} as the task of voice conversion (VC) in which the expressivity of the source speaker is preserved during conversion while the identity of a target speaker is transferred. To do so, an original neural-VC architecture is proposed based on sequence-to-sequence voice conversion (S2S-VC) in which the speech prosody of the source speaker is preserved during conversion. First, the S2S-VC architecture is modified so as to synchronize the converted speech with the source speech by mean of phonetic duration encoding; second, the decoder is conditioned on the desired sequence of F0-values and an explicit F0-loss is formulated between the F0 of the source speaker and the one of the converted speech. Besides, an adversarial learning of conversions is integrated within the S2S-VC architecture so as to exploit both advantages of reconstruction of original speech and converted speech with manipulated attributes during training and then reducing the inconsistency between training and conversion. An experimental evaluation on the VCTK speech database shows that the speech prosody can be efficiently preserved during conversion, and that the proposed adversarial learning consistently improves the conversion and the naturalness of the reenacted speech.

\end{abstract}

\vspace{0.1cm}

\vspace{-0.25cm}

\section{Introduction}
\label{sec:introduction}

\subsection{Context and Related works}

Voice conversion (VC) consists of digitally altering the voice of an individual - e.g., its identity, accent, or emotion - while maintaining its linguistic content unchanged. Primarily applied to identity conversion, VC has considerably gained in popularity and in quality thanks to the advances accomplished with neural VC (see for instance, \cite{Tod16, Lor18, zhao2020voice}).  Neural VC is now widely considered as a standard in VC and has reached a highly-realistic rendering of voice identity conversion from a small amount of data of a target voice.
Similarly to face manipulation, voice conversion has a wide range of potential applications, such as voice cloning and deep fake in the fields of entertainment and fraud \cite{lorenzotrueba2018steal}, anonymization of voice identity in the field of security and data privacy \cite{Vincent20, ericsson2020adversarial}, or digital voice prosthesis of impaired speech in the field of digital healthcare \cite{Wan20}.

%
%
Introduced in the late 2000's, neural VC has gradually moved towards many-to-many and non-parallel datasets allowing the scalability of neural VC to large and multiple speakers datasets with the assumption that the increase of data will induce a substantial increase in terms of quality and naturalness of the VC. In particular, starGAN-VC \cite{kameoka_stargan-vc:_2018, kaneko_stargan-vc2:_2019} has been proposed to extend the paradigm of cycle-GAN to many-to-many and non-parallel VC by proposing a conditional encoder-decoder architecture. As opposed to the cycleGAN-VC, starGAN-VC is composed of a single encoder-decoder in which the decoder is conditioned on the speaker identity to be converted. In the starGAN-VC, a discriminator is employed to distinguish between real original speech and fake converted speech together with a speaker classifier is employed to determine whether the converted speech has been produced by the target speaker. As in the cycleGAN, a cycle-consistency loss attempts to preserve the linguistic content during conversion. 
To overcome the issue of linguistic preservation trough conversion, phonetic posterior-grams (PPGs) \cite{sun_phonetic_2016, mohammadi2019one} have been proposed to integrate explicitly linguistic content information by mean of time-aligned phonetic posterior probabilities that are used to condition the conversion. 
\vspace{-0.25cm}

More recently, neural VC architectures has reformulated the VC problem as a simple auto-encoder \cite{Lu2019, qian2019autovc, Zhang_2020} (further referred to as AE-VC). The underlying idea is to provide a more structured representation of the information in the process of VC, in particular by explicitly encoding and disentangling linguistic content. VC is then achieved by manipulating the speaker identity by conditioning the decoder on the desired speaker identity. This has been achieved either by carefully tuning a bottleneck to encode only the linguistic content information as in \cite{qian2019autovc} or through adversarial learning of linguistic content and speaker identity codes as in \cite{Zhang_2020}. The advantage of those architectures is that the ground truth is known during training, so that a simple reconstruction loss can be efficiently applied (as opposed to the starGAN-VC). Moreover, linguistic and speaker information are explicitly represented and learned through disentanglement, which should idealistically provide a more accurate control during conversion.  Those architectures have opened the possibility of processing VC from a very small number of examples of the target speaker (at the extreme from one-shot \cite{Lu2019} or zero-shot \cite{qian2019autovc}).
One important assumption of these architectures relies on the fact that though a simple reconstruction is used during training (thus no conversion is actually learned), the  manipulation of the speaker identity is still effective during conversion. However, this might seem somehow inconsistent since the actual conversion is not used during training (as opposed to the starGAN-VC).

\subsection{Limitations and Contributions}

Though VC systems can now achieve highly-realistic voice identity conversion \cite{Lor18, zhao2020voice}, the only information which is assumed to be preserved from the source speaker through conversion is the actual linguistic content of the spoken utterance (i.e., the text transcript). This constitutes an important limitation of current VC systems since one could desire to preserve some aspects of the source speaker through conversion, e.g. the prosody and the expressivity of the source speaker. By direct analogy with the image domain, face reenactement is defined by preserving the expression of the source face while converting the identity to the one of a target face \cite{Thi16}. This provides a more flexible and controllable transformation as it is commonly known as deep fakes, in which the source face provides the desired pose and expression to be rendered with the target identity. 

The same definition can be applied in voice conversion, which will be further referred to as {\em voice reenactement}. Voice reenactement aims at preserving some aspects of the prosody of the source speaker during conversion, namely its timing - by means of phoneme durations - and its fundamental frequency (F0) both being commonly considered as the most important aspects of speech prosody. Though timing is ineherently preserved in some VC architectures such as cycleGAN-VC or starGAN-VC, this is not the case for all architectures including some of the aforementioned VC architectures based on auto-encoders. Besides, the only research to date  to the knowledge of the authors on the F0-preservation of the source speaker in VC has been proposed by \cite{9054734}, which is solely obtained by conditioning the decoder of Auto-VC on the desired sequence of F0-values. 
Moreover, AE-based VC suffers from the fact that the conversion is actually not learned during training while the GAN-based VC suffers from the fact that no ground truth is available to learn the conversion (and instead applying the adversarial loss). However, research on face manipulation including the original starGAN \cite{Choi18} have presented strategies that can make both advantage of reconstruction of the original real data and adversarial learning of the manipulated data \cite{He19}. 
The main contributions of this paper can be listed as follows: 
\begin{enumerate}
    \item The task of {\em voice reenactement} is introduced with the objective of preserving the expressivity of the source speaker during conversion while transferring the identity of the target speaker;
    \item An original solution to this task is presented by  preserving the timing and the F0 of the source speaker during conversion within a S2S AE-VC architecture \cite{Zhang_2020}. In particular, a F0-loss is explicitly formulated and exploited during training; 
    \item Inspired by the original formulation of the starGAN, an adversarial learning of identity conversion is integrated within the AE-VC architecture so as to both take advantage of reconstruction and conversion during training and reducing the inconsistency between training and conversion. 
\end{enumerate}

\section{Proposed Method}
\label{sec:methodology}

\subsection{Original S2S Neural VC}
\label{sec:S2SVC_orig}

\begin{figure*}[!ht]
\vspace{-0.5cm}
\begin{center}
\hspace{-1cm}\includegraphics[height=6.5cm]{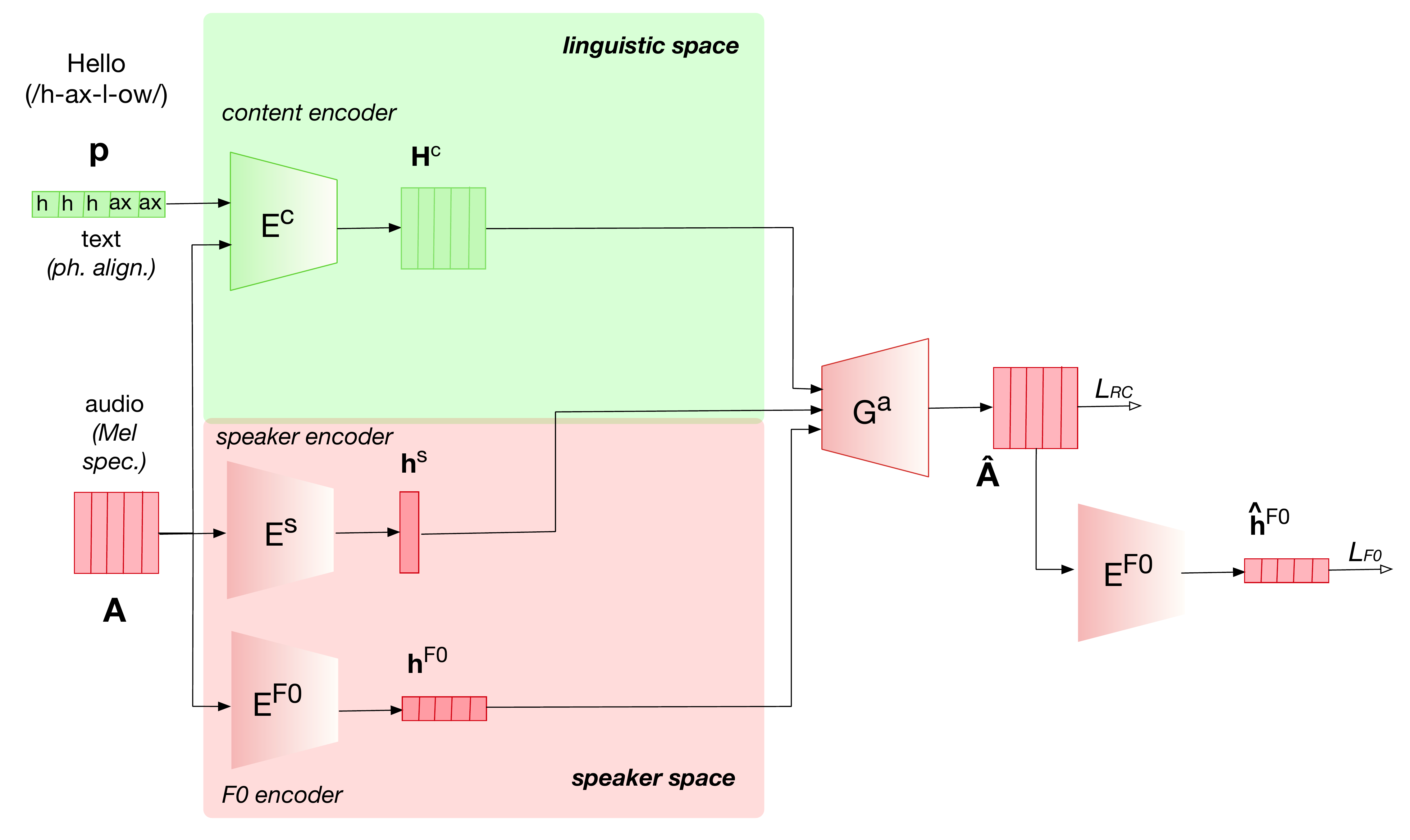}
\hspace{-0.25cm} \includegraphics[width=0.93\columnwidth]{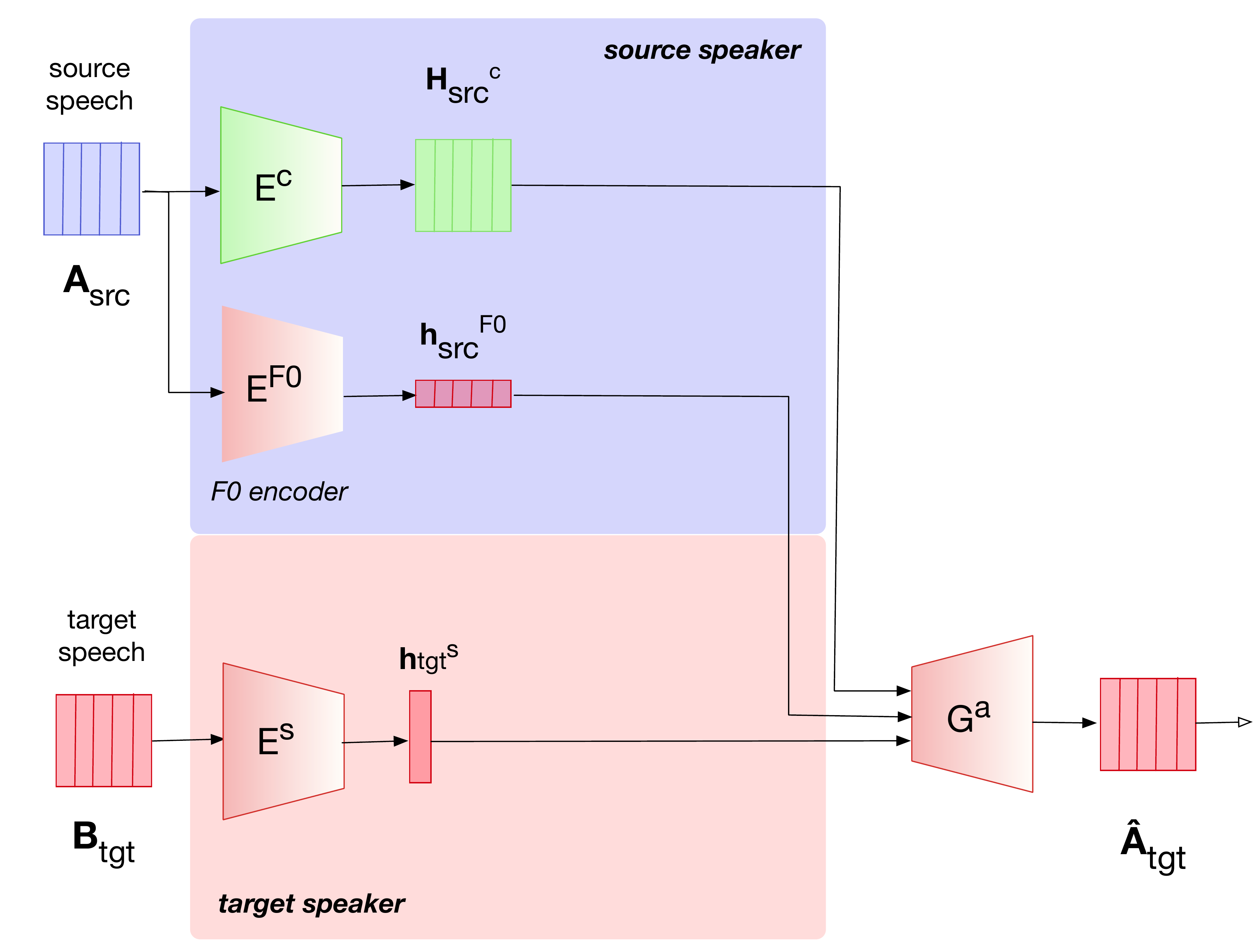}
\caption{Architecture of the proposed S2S-VC system. On left: The original VC comprises linguistic and speaker encoder $E^c$ and $E^s$, and a decoder $G^a$ which reconstructs the speech signal $\widehat{\mathbf{A}}$ conditioned on content embedding $\mathbf{H}^c$ and speaker embedding $\mathbf{h}^s$. A F0 encoder $E^{F_0}$ is added to encode the sequence of F0 values from the original Mel-spectrogram, and a F0 loss is computed between the F0 of the reconstructed speech signal and the one of the original speech signal. On right: During conversion, content embedding $\mathbf{H}^c_{src}$ and F0 $\mathbf{h}^{F_0}_{src}$ is extracted from speech signal of the source speaker and the speaker embedding $\mathbf{h}^s_{tgt}$ of the target speaker.} 
\label{figure:S2SVC_orig}
\end{center}
\vspace{-0.5cm}
\end{figure*}

The VC framework used in this paper is rooted on a sequence-to-sequence (S2S) auto-encoder as proposed in \cite{Zhang_2020}, in which disentangled linguistic and speaker representation are encoded through dedicated encoders as illustrated in Figure \ref{figure:S2SVC_orig}.
The inputs of the system are the speech signal matrix $\mathbf{A}$ represented by the Mel-spectrogram computed on $T$ time frames, and  the  sequence of $T$ phonemes $\mathbf{p}$ corresponding to the  phonetic transcription of the input text aligned to the corresponding speech signal. 
Dual encoders $E^c$ and $E^s$ are employed to encode linguistic content and speaker information. The speaker encoder $E^s$ converts the speech signal $\mathbf{A}$ into a time-independent vector $\mathbf{h}^s$, since it is assumed that the identity of a speaker does not vary within an utterance. A speaker classification loss $L_{SE}$ is defined between the speaker identity predicted from $\mathbf{h}^s$ and the true speaker identity $\mathbf{s}$. The linguistic encoder $E^c$ converts either the phoneme sequence $\mathbf{p}$ or the speech signal $\mathbf{A}$  into a shared linguistic embedding $\mathbf{H}^c$ through a contrastive loss. Contrary to \cite{Zhang_2020}, the linguistic embedding has the same length $T$ as the aligned phoneme sequence (and the Mel-spectrogram), so that the time information is preserved during encoding.
A S2S decoder $G^a$ conditioned on the content embedding $\mathbf{H}^c$ and the speaker embedding $\mathbf{h}^s$ is employed to reconstruct an approximation $\widehat{\mathbf{A}}$ of the original speech signal $\mathbf{A}$. A reconstruction loss $L_{RC}$ is defined between the reconstructed speech signal $\widehat{\mathbf{A}}$ and the original speech signal ${\mathbf{A}}$.
During training, the S2S-VC neural network is pre-trained on a multiple-speakers dataset, and then fine-tuned with respect to a given pair of source and target speakers. During conversion, the recognition encoder $E^c$ computes the content embedding $\mathbf{H}^c_{src}$ corresponding to one utterance ${\mathbf{A}_{src}}$ of the source speaker; and the speaker encoder $E^s$ computes the speaker embedding $\mathbf{h}^s_{tgt}$ corresponding the one utterance ${\mathbf{A}_{tgt}}$ of the target speaker. Then, decoder $G_a$ is conditioned on the linguistic embedding $\mathbf{H}^c_{src}$ and the speaker embedding $\mathbf{h}^s_{tgt}$ to generate the utterance ${\mathbf{B}_{tgt}}$ with the identity of the target speaker. Without explicit notification, the architecture and parameters of the S2S-VC are the same as those described in \cite{Zhang_2020}.

\subsection{Pre-processing and Post-Net}
The S2S-VC operates on a Mel-spectrogram representation of the speech signal. For the signal analysis we follow the parameterization proposed in \cite{liu_wavenet_2018}, that is the input signal is down-sampled to 16kHz, converted into an STFT using an Hanning window of $50$ms with hop size of $12.5$ms and an FFT size of $2048$. We then use $80$ Mel bins covering the frequency band from $0$ to $8$Khz and convert the result into log amplitude domain. 
A standardization of the log Mel-spectrogram is applied at the input of the VC system, i.e. on each Mel bin, removing the mean and diving by the standard deviation, which are pre-computed on the entire training dataset.
For rendering the speech signal from a generated Mel-spectrogram, a multi-speaker neural decoder  \cite{Roe22} was employed. This decoder was trained over $900,000$ iterations using all samples of the VCTK database with batch size $50$, and segment length $375$ms using the Adam optimizer with learning rate $1e^{-4}$.

\subsection{Proposed S2S Neural Voice Reenactemen}

\subsubsection{Contribution 1: S2S-VC with time-synchronization and F0-conditioning}

S2S-VC architecture does not ensure time-synchronicity between the original speech and the converted speech, by definition of the sequence-to-sequence architecture. The S2S-VS is basically composed on a speech-to-text encoding time-compression and a text-to-speech decoding time-decompression with a similar architecture as the Tacotron (\cite{tacotron2017, tacotron2_2018}). 
\textcolor{black}{In order to preserve time-synchronization between the original and the reconstructed speech signals, the time dimension of length $T$ is preserved all through the network, from the original speech signal ${\mathbf{A}}$ to the linguistic embedding $\mathbf{H}^t$, and to the reconstructed speech signal $\widehat{\mathbf{A}}$.
To do so, the auto-regressive S2S part of the recognition encoder $E^t$ and the  decoder $G^a$ are modified accordingly by employing simple recurrent architectures.
The recognition encoder $E^t$ is composed of two bidirectional LSTM layers of dimension 128  followed by a fully connected layer (FC) of dimension 128, resulting in a linguistic embedding of dimension ($128 \times T$).
The decoder $G^a$ is using two bidirectional LSTMs of dimension 128 each and a Fully Connected layer of dimension 80 which outputs an approximated Mel-spectrogram with the same dimensions as the input Mel spectrogram, i.e., ($80 \times T$). 
These simplifications enable time-synchronous conversions and a consequent saving in computational time: approximately 33\% of the computational time for training on our server with a single GPU.}
In order to preserve the F0 of the original speech signal during conversion, a F0 encoder $E^{F_0}$ is added and a F0-loss is explicitly formulated (as opposed to \cite{9054734}). The F0 encoder converts the input Mel-spectrogram into a corresponding sequence of F0 values. This value corresponds to the estimated $F0$ value for voiced frames and to $0$ for unvoiced frames.  
This encoding is passed to condition the decoder $G_a$ in addition to the linguistic  and speaker embeddings to create the Mel-spectrogram. A F0 loss is defined as the mean square error between the F0 values of the generated speech and the F0 values of the original speech, as
\begin{equation}
    \mathcal{L}_{F_0}(\mathbf{h}^{F0}, \mathbf{h}^{F0}) = \frac{1}{T} \sum_{t=1}^T (\mathbf{h}^{F0}(t) - \widehat{\mathbf{h}}^{F0}(t))^2
\end{equation}
The F0 encoder $E^{F_0}$ is described in \cite{Ard19}, pre-trained on the training set of the VCTK speech database, and then fixed during VC training, i.e. is only used to compute the F0 loss $\mathcal{L}_{F_0}$. Contrary to \cite{9054734} this loss explicitly constrains the $F0$ to the desired $F0$ by defining a dedicated loss. 
This ensures that the $F0$ of the converted speech to be effectively preserved during conversion. 
This loss is added to the reconstruction loss with a weight $\lambda_{F0}$ varying linearly from $10e^{-6}$ to $10e^{-2}$ with the effect of increasing gradually the importance of the F0 preservation versus the reconstruction loss during training.
During conversion, the $F0$ can be transferred from an utterance of a source speaker or fixed arbitrarily (e.g., by applying transposition or setting any arbitrary values). In this paper, the $F0$ used for conditioning was adapted to the range of the target speaker in order to prevent unnatural converted speech that would be caused by some important difference between the respective ranges of the source and target speakers (typically when converting a male to a female or conversely). This was accomplished by normalizing the $F0$ values corresponding to the sentence of the source speaker with respect to the log(F0) mean and standard deviation of the  target speaker.  \\

\subsubsection{Contribution 2: Adversarial Training of S2S-VC}

S2S-VC basically relies on an auto-encoder optimizing a reconstruction loss between the original speech signal and the reconstructed one. This is somehow inconsistent with the conversion in which the identity of the speaker is manipulated during conversion (eventually, the F0 and timing) . In the case of a conversion, one does not have access to the ground truth speech signal and thus one cannot apply the reconstruction loss of the auto-encoder. To overcome this limitation and construct a VC system whose training is more consistent with conversion, we propose to split the training process into two modes, as inspired by the original starGAN \cite{Choi18}: a {\em reconstruction} loss corresponding to the classical auto-encoder in which the reconstruction loss can be computed; an {\em adversarial} loss, in which we assume that the true speech signal may not be available. This is typically the case in which least one of the codes conditioning the decoder is manipulated. For this mode, we introduce an adversarial module which is similar to the one used in a GAN. A discriminator $D^{adv}$ is optimized to distinguish between the real speech samples and the converted ones, while the decoder $G^a$ is optimized to fool the discriminator. During training, each samples contained in a batch is both passed to the decoder in the reconstruction mode with original unchanged codes and in the conversion mode with unchanged or manipulated codes. In this paper, only the speaker identity is manipulated so that the reconstruction has the right identity and the conversion mode a randomly picked  identity. The total loss $ \mathcal{L}_{GEN}$ including reconstruction and conversion losses can then be expressed as,
\begin{equation}
    \mathcal{L}_{GEN} = \mathcal{L}_{RC} + \lambda_{adv} \mathcal{L}_{ADV}
\end{equation}
For clarity, the F0 loss is not specified in the above loss but is still used as previously described.
In this paper, the discriminator $D^{adv}$ is composed of $4$ convolution layers with $128$ filters of size $(3\times3)$ with a stride of $(2\times2)$ and $\lambda_{adv}=1$ was found to provide a good trade-off between the effect of the reconstruction and the conversion on the decoder $G^a$.

\begin{table}
    \caption{Root mean square error of the reconstructed F0 (in Hz.).}
\footnotesize
    \centering
    \begin{tabular}{l|c|c|c|c}
    VC system               & M-to-M & F-to-F & M-to-F & F-to-M \\
    \hline
    \hline    
    F0 cond w/ adv. same id & 2.712        & 5.686            & 2.970          & 5.090 \\
    F0 cond w/ adv. diff id & 2.574        & 6.246            & 4.171          & 5.252 \\
    \end{tabular}
    \label{tab:F0}
    \vspace{-0.35cm}
\end{table}

\begin{table*}[!]
\vspace{-0.75cm}
\caption {\label{tab:res} MOS obtained for the different VC systems.}
\footnotesize
\centering
\begin{tabular}{l|l|l|l|l|l|l}
\hline
VC system & \multicolumn{2}{c}{Male-to-Male} & \multicolumn{2}{c}{Female-to-Female} & \multicolumn{2}{c}{TOTAL} \\
                & Similarity   & Naturalness & Similarity   & Naturalness & Similarity   & Naturalness\\ \hline\hline
Orig: target speaker                         & 4.92 & 4.94  & 4.98 & 4.97 & 4.98 & 4.96  \\ \hline
F0 cond.                         & 3.90 & \textbf{3.38}   & 3.93   & 2.85  & 3.92  & 3.09   \\ \hline
F0 cond w/ adv. same id                        & 3.90 & 3.15 & 3.94 & 2.91  & 3.96 & 3.14\\ 
\hline
F0 cond w/ adv. diff id                        & \textbf{3.91} & 3.16 & \textbf{4.23} & \textbf{3.21} & \textbf{4.06} & \textbf{3.18} \\ \hline
 \hline
\end{tabular}
\vspace{-0.25cm}
\end{table*}

\section{Experiments}
\label{sec:experiments}

\subsection{Speech dataset}

The English multi-speaker corpus VCTK 
\cite{vctk2017}  is used for training and testing.  The VCTK dataset contains speech data uttered by 110 speakers and the corresponding text transcripts. Each speaker read about $400$ sentences selected from English newspaper, which represents a total of about $44$ hours of speech. 
Four speakers
All speakers are included into the training and validation sets. For each speaker, we split the database in a training set with 90\% of the sentences and a validation set with 10\% of them.  The  total  duration  of the database is around $27$ hours after removing silences at the beginning and at the end of each sentence.

\subsection{Subjective Experiment Setup}

The experiment consisted into the judgment by listeners of: the similarity of the converted speech to the target speaker and the naturalness of the converted speech, using 5-degree MOS scale as commonly used for the experimental evaluation of VC algorithms. 
Each participant had to judge 15 speech samples which were randomly selected among the total number of speech samples produced for the subjective experiments. The experiment was conducted on-line. Participants were encourage to perform the experiment in a quiet environment and with a headphone. 
Four speakers were used for the experiment: two males (p232 and p274) and two females (p253 and p300) with eight randomly chosen sentences per speaker in the validation set. Conversion were computed between male speakers and between females speakers, resulting in two male and two female conversion setups.
Four configurations were compared: 1) the original audio signal and the converted speech with~: 2) time-synchronous and F0-preserved VC system (referred to as F0 cond.); 3) time-synchronous and F0-preserved VC system with discriminator trained only with the true speaker identity (same identity as the source speech utterance) , referred to as F0 cond. w/adv same id; 4)  time-synchronous and F0-preserved VC system with discriminator trained only with varying speaker identities (different identities from the one of the source speech utterance), referred to as F0 cond. w/adv diff id \footnote{Converted speech samples are available at: \url{http://recherche.ircam.fr/anasyn/obin/VC_EUSIPCO22.html}}.

\subsection{Results and Discussion}

Table \ref{tab:F0} presents the root mean squared error of the F0 of the converted speech with the proposed configurations for the 4 speakers used in the experiment. The error is around 5 Hz. in average which is not audible in most cases. This shows the efficiency of the proposed F0 preservation strategy, even in combination with the adversarial loss.
Table \ref{tab:res} presents the mean MOS scores of the compared system configurations obtained with 25 participants. On the one hand, the baseline VC system using time-synchronization and F0 preservation presents fair to good similarity to the target speaker (MOS=3.92 for similarity), even though the timing and the F0 contours is inherited from the source speaker. Also, the naturalness is also rated to the fair range (MOS=3.09). One can observe that the naturalness is degraded with the female speakers (MOS=2.85) as compared to the male speakers (MOS=3.38). These results are consistent with the ones obtained in the literature about S2S-VC, as in \cite{Zhang_2020}. This indicates that the application of a timing and a F0 contour from a different speaker does not degrade consistently neither the similarity nor the naturalness of the converted speech.
On the other hand, the proposed VC system with time and F0 preservation together with adversarial loss on varying speaker identities improves the scores in almost all cases compared to the baseline. The overall similarity to the target speaker is good (MOS=4.06) and the naturalness of the conversion is fair (3.18). The improvement in similarity is particularly substantial for the female speakers (MOS=4.23) while in the same time the difference in naturalness between the male and female conversions is much less pronounced (MOS=3.16 for male speakers and MOS=3.21 for female speakers). This indicates that the add of the discriminator not only helps to improve the naturalness of the converted speech (by suppressing perceptible artifacts) but also increases the similarity to the target speaker. Also, the use of varying speaker identities with the adversarial loss improves the scores in all cases compared to using only the true identity of the speaker. This is probably mainly due to the fact that the discriminator is being more efficient when subject to a larger variety of sentence and speaker identities.

\vspace{-0.15cm}

\section{Conclusion}
\label{sec:conclusions}

This paper presented a S2S-VC algorithm which allows to preserve the timing and the F0 of the source speaker during conversion. Moreover, an adversarial module is added so that the S2S-VC can learn both from real speech samples as well as manipulated ones. Experimental evaluation on the VCTK speech database shown that the F0 is effectively preserved during conversion and that the adversarial module clearly helps to reduce the audible artifacts of the conversion as well as improve the perceived identity of the converted speech. Further research will investigate the manipulation of timing and F0 during training, in addition to the speaker identity.

\vspace{-0.15cm}

\section{Acknowledgements}
\label{sec:acknowledgements}
The research in this paper has been funded by the ANR project TheVoice :
ANR-17-CE23-0025. 

\bibliographystyle{IEEEbib}
\bibliography{refs}

\end{document}